# Is the kinetic resonance in human glucokinase genuine?

*Leonid N. Christophorov**

Bogolyubov Institute for Theoretical Physics, NAS Ukraine. 14b Metrologichna Str., 03143 Kyiv, Ukraine

**ABSTRACT**. In the recent paper by Mu *et al* (J. Phys. Chem. Lett. 2021, 12, 2900) the maximal cooperativity of human glucokinase is explained by the "kinetic resonance" effect derived from a minimal three-state model. However, a closer inspection of the latter shows that this effect seems to be rather a particular quantitative coincidence than a general phenomenon that reflects a physical mechanism of conformational regulation.

**TOC GRAPHICS**

$$k_{cat} \stackrel{?}{\approx} k_{ex}$$

-------------------------------

* E-mail: lchrist@bitp.kiev.ua

Human glucokinase (GCK) is a known example of a monomeric enzyme exhibiting cooperativity (so-called allokairy[1-3]). In the recent Letter[4] its authors explain a pronounced manifestation of this cooperativity observed experimentally[2] by the "kinetic resonance" effect in a simplified three-state model. To the latter, they apply their previous theoretical conclusions for a somewhat different and more complex model.[5] As a result, the analysis becomes unclear and hides the physics of the phenomenon, there appear some inaccuracies, and the very kinetic resonance effect – at least, in the form it is declared – looks not too convincing. Meanwhile, the model under study can be fully analyzed in a standard way known long since.[6,7] With simpler designations (without multiple indices), this model is depicted in Figure 1.

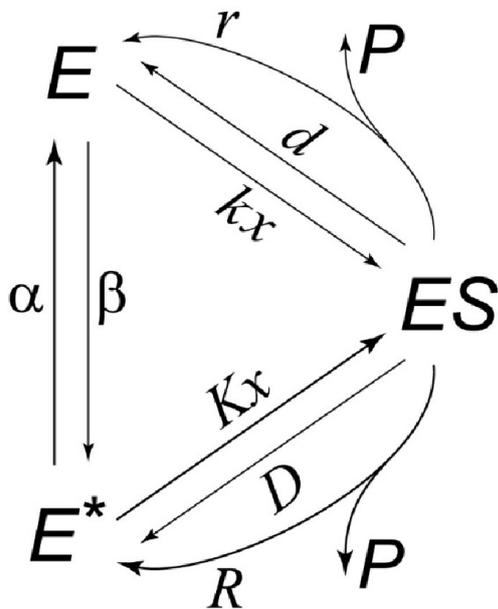

**Figure 1.** Conformationally splitted Michaelis-Menten scheme analyzed in ref 4. States $E$, $E^*$ represent those of unliganded GCK that differ in affinities to substrate $S$ (glucose); henceforth, its concentration $[S]$ is denoted as $x$. Here $Kx$, $kx$ are the rate constants of substrate binding, $D$, $d$ are those of unproductive dissociation, $R$, $r$ are those of catalytic conversion of $S$ into product $P$, and $\alpha$, $\beta$ are the rate constants of conformational interconversions.

Since the authors of ref 4 impose the conservation condition $P_E + P_{E^*} + P_{ES} = 1$ for the state probabilities, then the corresponding kinetic equations are in fact identical to those for concentrations in the standard chemical kinetics.[8] Solving them in a trivial way, for the stationary reaction velocity per enzyme molecule $v = (R+r)P_{ES} \equiv k_{cat} P_{ES}$, one has

$$\frac{v}{k_{cat}} = \frac{x^2 + Cx}{x^2 + x(B+C) + AC}, \qquad (1)$$

or, as it is presented in ref 4,

$$\frac{1}{v} = \frac{1}{k_{cat}} \left( 1 + \frac{A}{x} + \frac{B-A}{x+C} \right), \tag{2}$$

where $k_{cat} = R + r$, and

$$A = \frac{(\alpha + \beta)(D + d + R + r)}{\alpha k + \beta K},$$
$$B = \frac{K(d+r) + k(D+R)}{Kk}, \tag{3}$$
$$C = \frac{\alpha k + \beta K}{Kk}.$$

Referring to the previous work,[5] the authors assert that the presence of positive cooperativity (i.e., flexion of the curve $v(x)$) is determined by the inequality $A > B$. This is incorrect, however. The sigmoidicity of $v(x)$ requires the existence of a positive root to the equation $v''(x) = 0$. An example of the corresponding investigation of fractional polynomials like eq 1 was given by Ferdinand.[6] In our case, equaling the second derivative of eq 1 to zero, one has a cubic equation $Bx^3 + 3ACx^2 + 3AC^2 x + AC^2(B + C - A) = 0$. As $A, B, C > 0$, it has a positive root if only the last term is negative. This results in the condition $A > B + C$, or, in terms of the system parameters, in the inequality

$$Kk[\alpha(D+R) + \beta(d+r) - 2\alpha\beta] > K^2\beta(d+r+\beta) + k^2\alpha(D+R+\alpha). \tag{4}$$

Even under specific assumptions made in ref 4, to deduce the kinetic resonance condition $k_{cat} \approx k_{ex}$ (i.e., $R + r \approx \alpha + \beta$) of maximal cooperativity from eq 4 looks problematic. Inexplicably, the authors do not provide the model/real parameter values that secure the values $A$, $B$, $C$ used in their Fig. 2 illustrating the kinetic resonance.

Somewhat clearer for analysis is the case that is implied in all early explanations of kinetic cooperativity, starting from ref 7, see also the reviews in refs 9,10. Precisely, the key point for cooperativity is the presence of a *sufficiently slow* conformational transition $E^* \to E$ into *more* *stable* (but *of lower affinity*, $k < K$) state E. This asymmetry implies that $\alpha$, being sufficiently small, is nevertheless considerably greater than $\beta$; for clarity, let $\beta = 0$. Then the condition (4) degenerates into the following inequality:[11,12]

$$\alpha < \left(\frac{K}{k} - 1\right)(D + R), \tag{5}$$

which, indeed, requires certain slowness of the mentioned transition. Note, however, that in this case neither $A$, nor $B$ depend on $\alpha$, and $C = \alpha/K$. The kinetic resonance condition (which would be $\alpha \approx R + r$ here) hardly follows from eq 5, too. To evaluate the degree of cooperativity, one can try to use the Hill coefficient $n_H = d\log[v/(v_m - v)]/d\log x$, where $v_m = R + r$ in our case. In terms of $A$, $B$, $C$

$$n_H = \frac{Bx^2 + 2ACx + AC^2}{Bx^2 + (A+B)Cx + AC^2}. \tag{6}$$

It is easy to see that $n_H(x \to 0, \infty) \to 1$ but can exceed 1 in a certain domain of concentration $x$ and has its maximal value $n_{H\,\max} = \left(2A + 2A\sqrt{A/B}\right)/\left[2A + (A+B)\sqrt{A/B}\right] < 2$ at $x_{\max} = C\sqrt{A/B}$, if $A > B$. Note that the latter condition differs from the necessary condition for positive cooperativity $A > B + C$. In other words, the Hill coefficient (6) can exceed 1 while positive cooperativity is still absent (see curves *b* in Figure 2 below). Conventionally, the greater $n_{H\,\max}$, the more pronounced cooperativity, although one should bear in mind that concentration $x_{\max}$ required for the Hill coefficient noticeably greater than 1 can turn out to be rather high.

Turning back to the case $\beta = 0$, I have found a set of parameters that ensures the values $A = 125$, $B = 3$, $C = 0.4$ used in ref 4 for fitting the experimental curve of ref 2, see curves *a* in Figure 2. Note that all the rate constants in Figure 1 and eq 3 can be scaled by one and the same factor without changing the values of $A$, $B$, $C$ and the curves in Figure 2. At first glance, the resonance condition $k_{ex} \approx k_{cat}$ (i.e., $\alpha + \beta \approx R + r$) approximately holds. However, keeping this condition intact while introducing even a relatively weak backward conformational transition (say, $\alpha = 70, \beta = 10$) immediately destroys positive cooperativity, see curves $b$ in Figure 2. Adding redistribution of catalytic rates leads to even negative cooperativity (curves $c$). On the other hand, taking noticeably greater $R$ outside the resonance condition results in even better cooperativity (curves $e$). Moreover, violating this condition even stronger ($\alpha + \beta = 35$ while $R + r = 480$, curve $d$) practically reproduces the GCK cooperativity profile represented by curve $a$.

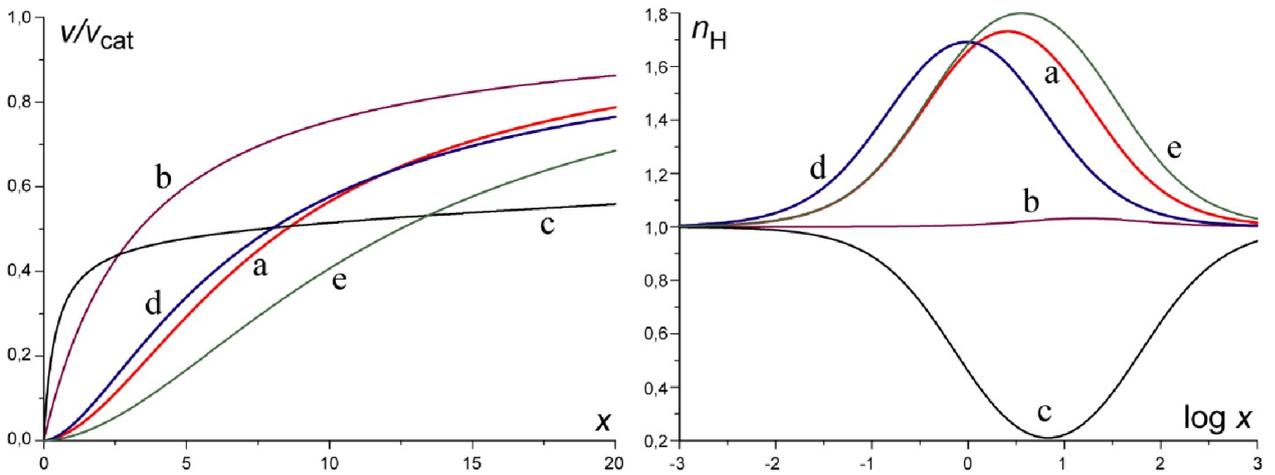

**Figure 2.** *Left*: The dependence of the reaction velocity on substrate concentration. (a) $K = 200$, $k = 0.7$, $D = 0.7$, $d = 1$, $R = 85$, $r = 0.8$, $\alpha = 80$, $\beta = 0$; (b-c) the same parameter values except $\alpha = 70$, $\beta = 10$ (curve b), or $\alpha = \beta = R = r = 40$ (curve c). (d) $K = 200$, $k = 4$, $D = 100$, $d = 8$, $R = 480$, $r = 0$, $\alpha = 35$, $\beta = 0$. (e) $K = 200$, $k = 0.7$, $D = 0.7$, $d = 1$, $R = 200$, $r = 0.8$, $\alpha = 80$, $\beta = 0$. *Right*: The corresponding dependence of the Hill coefficient.

One can therefore conclude that the kinetic resonance condition does not seem crucial for a pronounced positive cooperativity. The origin and degree of the latter is determined rather by the asymmetry of conformational interconversions between less stable state $E^*$ of high affinity and more stable state $E$ of low affinity, with relaxation $E^* \rightleftarrows E$ being sufficiently slow and sufficiently biased. It is these physical conditions that ensure the reaction pathway via a more effective channel under higher substrate concentrations and thereby positive kinetic cooperativity.[1,7,8,10-12] On the other hand, it remains unclear how the kinetic resonance effect claimed in ref 4 could be explained physically. It seems to be a particular quantitative coincidence under rather specific assumptions than a general phenomenon that reflects a physical mechanism inherent in this oversimplified (yet multi-parametric) scheme of monomeric cooperativity.

ACKNOWLEDGMENT

The work is performed within Project 0116U002067 of the NAS of Ukraine.